\documentclass[aip,twocolumn,letterpaper]{revtex4}

\usepackage{fullpage}

\usepackage{graphics}
\usepackage{epsfig}

\usepackage{xcolor}

\usepackage[margin=1.0in]{geometry}

\begin{document}
\title{Judgment of paradigms for magnetic reconnection in coronal loops}
\author{Allen H Boozer}
\affiliation{Columbia University, New York, NY  10027\\ ahb17@columbia.edu}

\begin{abstract}

The traditional paradigm for magnetic field lines changing connections ignores magnetic field line chaos and requires an extremely large current density, $j_{max}\propto R_m$, flowing in thin sheets of thickness $1/R_m$, where $R_m$ is the magnetic Reynolds number.  The time required for a general natural evolution to take a smooth magnetic field into such a state is rarely considered.  Natural evolutions generally cause magnetic field lines to become chaotic.   A fast change in field line connections then arises on the timescale defined by the evolution multiplied by a $\ln(R_m)$ factor, and the required maximum current density scales as $\ln(R_m)$.  Even when simulations support the new paradigm based on chaos, they have been interpreted as supporting the old.  How this could happen is an important example for plasma physics of Kuhn's statements about the acceptance of paradigm change and on Popper's views on the judgment of truth in science.

\end{abstract}

\date{\today} 
\maketitle


In 1962, Thomas Kuhn wrote \cite{Kuhn} what the Encyclopaedia Britannica described as \emph{one of the most influential works of history and philosophy written in the 20th century} \cite{Britannica:Kuhn}.  He discussed the importance of paradigms in science and how difficult it is for a scientific community to accept a change in paradigm.  The physics of  changes in the connections of magnetic field lines illustrates Kuhn's point in a problem of great importance in both natural and laboratory plasmas.

The traditional paradigm for changes in field line connections was clearly stated by Schindler, Hesse, and Birn in their paper on general magnetic reconnection \cite{Schindler:1988}.  To obtain changes in field line connections at a rate consistent with plasmas with large magnetic Reynolds numbers, $R_m \sim 10^8 \mbox{  to } 10^{20}$, an intense current density, $j_{max}\propto R_m$ must arise in sheets of thickness $\propto 1/R_m$.  Hundreds of papers have been written on ways such an extreme current density  can be maintained if it were initially present, but the way an arbitrary magnetic field could evolve into such a state is rarely considered.

Papers by a number authors emphasize that chaotic magnetic-field-line trajectories fundamentally change the paradigm of magnetic reconnection from that of Schindler et al.  Chaos enters the theory of turbulent magnetic reconnection, and this topic was reviewed \cite{Lazarian:2020rev} by Lazarian, Eyink, Jafari, Kowal, Li, Xu, and Vishniac in 2020.  Eric Priest has been associated with a large body of work on three-dimensional structures that tend to concentrate currents and thereby lead to enhanced reconnection \cite{Priest:2016}.  In particular, he is known for his work on quasi-separatrix layers, which are essentially regions of field line chaos.  Reid, Parnell, Hood, and Browning \cite{Reid:2020}, have simulated a case in which the footpoint motions of magnetic field lines do not directly make the lines chaotic but drive large-scale instabilities that do.

\begin{figure}
\centerline{ \includegraphics[width=2.0in]{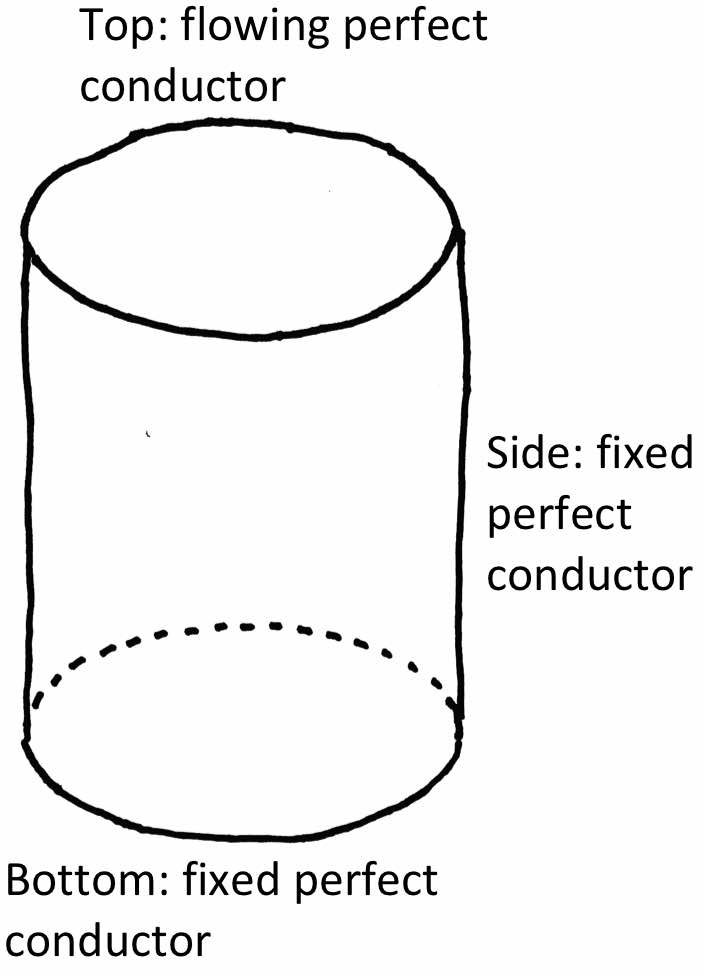}}
\caption{A perfectly conducting cylinder of height $L$ and radius $a$ encloses an ideal pressureless plasma.  All of the sides of the cylinder are fixed except the top, which flows with a specified velocity $\vec{v}_t$.  Initially, $\vec{B}=B_0\hat{z}$.  Each point $\vec{x}_b$ on the bottom of the cylinder defines a line of $\vec{B}$ that in an ideal evolution intercepts a specific point on the top $\vec{x}_t$ with   $\partial \vec{x}_t(\vec{x}_0,t)/\partial t= \vec{v}_t(\vec{x}_t,t)$ and $\vec{x}_0\equiv \vec{x}_t$ at $t=0$. The case of primary interest is when $\vec{v}_t$ is divergence free and chaotic.  This means the $2\times2$ Jacobian matrix  $\partial \vec{x}_t/\partial \vec{x}_0$ has a large singular value that increases exponentially in time and a small singular value that is the inverse of the large singular value.} 
\label{fig:cylinder}
\end{figure}

Allen Boozer has published a number of articles on the fundamental importance of chaos to magnetic reconnection and that show that the magnetic fields must have a non-trivial dependence on all three spatial dimensions for chaos to be important.  Two recent articles are \cite{Boozer:2021,Boozer:2022}.  He and Todd Elder also published  \cite{Boozer-Elder} a simple example, Figure \ref{fig:cylinder}, of corona-like footpoint motions directly driving magnetic field line chaos.   The changing of field line connections is rigorously defined in this model.  It is shown that going from spatially-constant straight magnetic field lines to a state in which a change in field line connections is unavoidable requires a time that is the ideal evolution time multiplied by $\ln(R_m)$ and the maximum current density need be only proportional to $\ln(R_m)$.  They found the currents lie in extremely contorted ribbons along the magnetic field lines, have a width proportional to $R_m$, and a thickness proportional to $1/R_m$.

It should be emphasized that here and in mathematics, chaotic trajectories do not mean random but that neighboring trajectories separate exponentially.  See the caption to Figure \ref{fig:cylinder}.

The central conclusions of the work by Boozer can be summarized in two sentences: \emph{Assume magnetic field line connections are well defined and non-ideal effects on the magnetic evolution are small.  Then, magnetic field lines go from a simple smooth form to having large and broadly-spread changes in their connections on a timescale that is approximately a factor of ten longer than the ideal evolution time when and only when the magnetic field lines become chaotic.}

It should be noted that although chaos exponentially enhances the breaking of magnetic field line connections, chaos can only spread, not destroy, the helicity injected by footpoint motion.  When the rate of helicity injection is greater than the large-scale resistive timescale of the plasma, flux-tube eruption must eventually occur \cite{Boozer-Elder}. This result would be false if helicity could be efficiently destroyed at small spatial scales.   Although it has been known starting in mid-1980's that this is not energetically possible \cite{Berger:1984,Thiffeault:2003}, destruction of helicity at small scales is commonly assumed in dynamo theory \cite{Rincon:2019}.

Karl Popper is widely acclaimed as one the twentieth century's most influential philosophers of science.  He famously stated \cite{Popper} that a scientific statement can never be proven to be correct but that it must in principle be testable.  The most reliable scientific statements have been tested and never proven false.

The central conclusions of Boozer's work on the changing of field line connections could be proven false in two ways:  (1) Find an evolving highly chaotic magnetic field that nonetheless preserves well-defined magnetic field line connections.  (2) Find an evolving non-chaotic magnetic field that nonetheless goes from being simple and smooth to large scale connection breaking on a timescale only an order of magnitude longer than the ideal evolution time, even when $R_m$ is of order $10^8$ or larger.

The changing of magnetic field line connections on a timescale approximately an order of magnitude longer than the evolution timescale,  even as $R_m\rightarrow\infty$, is a common feature of observations of natural plasmas and of laboratory experiments.  Having an appropriate paradigm for studying such of phenomena is clearly of major importance to the entire field of plasma physics.

Th difficulty of accepting a new paradigm is illustrated  by an August 15, 2022 arXiv posting by Yi-Min Huang and Amitava Bhattacharjee, \emph{Do chaotic field lines cause fast reconnection in coronal loops?} \cite{Huang-Bhattacharjee}.  They have carried out simulations of essentially the model of Figure \ref{fig:cylinder} and claim their simulations give a negative answer to the question in their title.   Their abstract explicitly lists eight papers by Boozer over the last decade and no papers by other authors except that of Boozer and Elder \cite{Boozer-Elder}.  The abstract says the Huang-Bhattacharjee simulations do not support the claims in these papers.  

Remarkably, the Huang-Bhattacharjee article finds that the field lines become chaotic and undergo large changes in their connections on a timescale approximately an order of magnitude longer than the ideal evolution time, independent of the resistivity; the changing of field line connections had become extremely fast at a time only 30\% later.  They did not mention and certainly gave no evidence for the existence of a comparison case in which field lines go from straight to changing connections on a timescale determined by the ideal evolution while remaining non-chaotic.

How can results that are seemingly consistent with a scientific statement be used to refute it? First, the authors say the changing of field line connections is not the fundamental definition of reconnection.  In their view, the fundamental definition is the development of the ``signatures of reconnection," such as intense current densities $j_{max}\propto1/\eta$ in thin sheets with $\eta$ the resistivity.  The development of such sheets is demonstrated by Figure 2 of their article.   

The Huang-Bhattacharjee ``signatures of reconnection" arise on a timescale only an order of magnitude longer than the ideal evolution timescale, so even using them as the definition of reconnection does not invalidate the two sentences of central conclusions.  The thin sheets of $j_{max}\propto1/\eta$ current densities are the primary requirement of the traditional Schindler et al paradigm for reconnection \cite{Schindler:1988}.   

The scaling of the maximum current density as $j_{max}\propto1/\eta$ in the simulations caused  Huang and Bhattacharjee to assume that the traditional paradigm remains an essential part of reconnection theory.  But, the $j_{max}\propto1/\eta$ in their simulations has another explanation---power balance.  The power put into the plasma by the moving top surface, $\vec{v}_t\cdot(\vec{j}\times\vec{B})$, can only be balanced by resistive dissipation, $\eta j^2$ integrated over the plasma volume, and then only when $j_{max}\propto1/\eta$.  Unlike Schindler et al reconnection, the chaos enhanced breaking of field line connections directly dissipates little energy.  The released energy must go into plasma kinetic energy, which means Alfv\'en waves.  Indeed as the changing of field line connections proceeds, the Huang-Bhattacharjee simulations see a rapid increase in the plasma flows.  An implication is that plasma viscosity $\nu_v$ could damp the input power as well as resistivity $\eta$.  However, in their simulations the Prandtl number, $P_r\equiv \nu_v \mu_0/\eta$ is unity and only $\sim 20\%$ of the input power is damped by viscosity.  When $P_r>>1$, large current densities are neither required to allow a change in field line connections nor to dissipate the input power. The velocity of the plasma will become smoother the larger the viscosity, but the timescale for the onset of large-scale changes in field line connections should have little dependence on viscosity when the chaos is directly driven by the boundary conditions as in the model of Figure \ref{fig:cylinder}. 

Huang and Bhattacharjee have agreed to do simulations with a large viscosity to determine whether sheets of intense currents persist.  However, a fundamental numerical problem complicates the interpretation of results.  Chaotic magnetic field lines by definition have a separation between infinitesimally separated pairs of lines that increases exponentially with distance along the lines.  The implication is that the numerical resolution required to preserve field line connections increases exponentially as the streamlines of the flow $\vec{v}_t$ exponentiate apart.  Any realizable numerical resolution becomes inadequate after a time that depends only logarithmically on the resolution when the resistivity is small.  It has been suggested to Huang and Bhattacharjee that they set $\eta=0$ in their code and keep only the viscosity $\nu_v$ and study the results for various values of the fluid Reynold number, $R_f=v_tL/\nu_v$.  Thought is required on how current densities are calculated when the numerical resolution is inadequate.  The numerical resolution issue could be solved by making the resistivity sufficiently large, which would require the magnetic Reynolds number $R_m$ be significantly smaller than than the ratio of the grid scale to the spatial scale of $\vec{v}_t$.  The applicability of results from relative small values of $R_m$ to physically important values, $R_m \sim 10^8 \mbox{  to } 10^{20}$, requires careful thought.

Essentially all natural flows are chaotic, which is why a radiator can heat a room in of order 10 min instead of the weeks expected from thermal diffusion.  The explanation is essentially the same for the rapid spread of the temperature and the rapid spread of magnetic field line connections \cite{Boozer:2021}.  Without diffusion the temperature in a room would evolve as $\partial T/\partial t+\vec{v}\cdot\vec{\nabla}T=0$ with $\vec{\nabla}\cdot\vec{v}=0$. The implications are that the constant temperature contours cannot break and that they enclose constant volumes.  But, a chaotic flow $\vec{v}$ produces exponentially increasing distortions in the shape of the constant-$T$ surfaces, which allow an exponentially small thermal diffusivity to spread the temperature.  The curves that define tubes of magnetic flux are distorted in an ideal evolution just as constant-$T$ contours are.  The exponentially increasing distortions allow resistive diffusion $\eta/\mu_0$ to diffuse field lines between the tubes.  Exponentially enhanced diffusion can occur for a scalar quantity, such as the temperature, in two dimensions, but three dimensions are required for the analogous effect for the vector $\vec{B}$  \cite{Boozer:2021}.   

A large Prandtl number is required not only to distinguish physics effects but also to be characteristic of space plasmas.  The values of $P_r$ given by Aschwanden \cite{Aschwanden} are many orders of magnitude larger than unity.  One reason is obvious.  The calculation of the cross-field viscosity in a plasma closely follows that of the ion thermal conductivity.  Small-scale turbulence that affects any part of the ion distribution function in velocity space can produce an enhancement.  Even in the relatively quiescent conditions of tokamak fusion plasmas, ion thermal diffusivities are of order the gyro-Bohm coefficient $D_{gB}\approx (\rho_i/a)D_{B}$, where the Bohm coefficient, $D_B\approx T/eB$ represents the largest microturbulent transport possible, $\rho_i$ is the ion gyroradius, and $a$ is the spatial scale of gradients.  Small-scale turbulence generally has little effect on the parallel resistivity $\eta_{||}$, which allows changes field line connections, because if any part of the electron velocity space can efficiently carry the current then it will.  Large enhancements in $\eta_{||}$ are not observed in tokamak plasmas.

One of the confusions in the literature is what is the definition of reconnection: a topology change in the magnetic field lines or an energy change in the magnetic field.  The two definitions are not equivalent.  Tokamak disruptions are spectacular changes in field-line topology.  Most of the magnetic surfaces can be destroyed and the current profile flattened on a timescale of milliseconds, when the naively expected timescale would be minutes, but the change in the poloidal field energy is $\sim$7.5\% with the poloidal field energy only a few percent of the magnetic field energy in the plasma.  Ideal instabilities can release magnetic energy on an Alfv\'enic time scale with no change in field line connections.  In the toroidal plasma physics community, the definition of reconnection is a topology change, which follows Parker and Krook's 1956 definition \cite{Parker-Krook:1956}.

The energy definition of reconnection tends to be more popular for reconnection in natural plasmas.  A major reason is that energy changes can be locally defined and the boundary conditions used for plasma problems that arise in nature are frequently too imprecise \cite{Boozer:2012a}  to make the field line changes well defined.

Connection changes are topological changes.  As is generally true with topological changes, where along the line the change took place is not necessarily well defined, nor is the speed of the change necessarily bounded.  What is bounded is the speed with which physical effects associated with the change take place.

These statements are easier to understand in the context of a tokamak disruption in which magnetic field lines go from lying in nested toroidal surfaces to chaotically covering the plasma volume.  At some point in time the last magnetic surface breaks and field lines that were bounded by that surface before that instant in time change at that time to trajectories that go to the surrounding walls.  Although the change in connections was in some sense instantaneous, the effects have various speeds: $j_{||}/B$ flattens on a shear Alfv\`en wave timescale and the electron energy spreads either by parallel thermal conduction or ballistically when the mean free path is long.

If plasma physics topics that are of great societal importance, such as space weather and magnetic fusion energy, are to be successfully addressed, optimal research paradigms must be developed and adopted.  Karl Popper and Thomas Kuhn have given clear statements about how paradigms are to be judged and the difficulties of having improved paradigms adopted.  The Huang-Bhattacharjee August 15 arXiv posting \cite{Huang-Bhattacharjee} offers lessons in this regard because their simulations do not support the point they claim.   For the lessons to be learned, the community needs access to information about the divergence between the implications of the results and the points made by the authors.

The Huang-Bhattacharjee simulations are important and should be published.  Their  simulation capabilities could greatly enhance the understanding of reconnection by the use of a physically relevant viscosity and by studying the eruption of coronal loops due to helicity accumulation.

 
\section*{Acknowledgements}
This work was supported by the U.S. Department of Energy, Office of Science, Office of Fusion Energy Sciences under Award Numbers DE-FG02-95ER54333, DE-FG02-03ER54696, DE-SC0018424, and DE-SC0019479.

\section*{Data availability statement}

Data sharing is not applicable to this article as no new data were created or analyzed in this study.


\end{document}